\documentclass[a4paper,11pt]{article}
\pdfoutput=1
\usepackage{jheppub}
\usepackage[noconfig]{refstyle}
\usepackage{longtable,tikz}
\usetikzlibrary{shapes,arrows}
\tikzstyle{block} = [rectangle, draw, text width=7em, text centered, rounded corners, minimum height=3em]

\let\eqref=\relax
\newref{eq}{name={eq.~},Name={Eq.~},names={eqs.~},Names={Eqs.~},rngtxt={-},refcmd=(\ref{#1})}
\newref{tab}{name={table~},Name={Table~},names={tables~},Names={Tables~}}
\newref{sec}{name={section~},Name={Section~},names={sections~},Names={Sections~}}
\newref{fig}{name={figure~},Name={Figure~},names={figures~},Names={Figures~}}
\numberwithin{equation}{section}

\newcommand{\be}{\begin{equation}}
\newcommand{\ee}{\end{equation}}
\newcommand{\field}[1]{\mathbb{#1}}
\newcommand{\oneon}[1]{\frac{1}{#1}}
\newcommand{\CP}{\field{P}}
\newcommand{\al}{\alpha}
\newcommand{\dd}{{\rm d}}

\title{All Complete Intersection Calabi-Yau Four-Folds}

\author[a]{James~Gray,}
\author[b]{Alexander~S.~Haupt}
\author[c]{and Andre~Lukas}

\affiliation[a]{Arnold-Sommerfeld-Center for Theoretical Physics,\\
        Department f\"ur Physik, Ludwig-Maximilians-Universit\"at M\"unchen,\\
        Theresienstra\ss{}e 37, 80333 M\"unchen, Germany}
\affiliation[b]{Institut f\"ur Theoretische Physik, Leibniz Universit\"at Hannover,\\
        Appelstra\ss{}e 2, 30167 Hannover, Germany}
\affiliation[c]{Rudolf Peierls Centre for Theoretical Physics, Oxford University,\\
        1 Keble Road, Oxford, OX1 3NP, U.K.}

\emailAdd{james.gray@physik.uni-muenchen.de}
\emailAdd{alexander.haupt@itp.uni-hannover.de}
\emailAdd{lukas@physics.ox.ac.uk}

\abstract{We present an exhaustive, constructive, classification of the Calabi-Yau four-folds which can be described as complete intersections in products of projective spaces. A comprehensive list of 921,497 configuration matrices which represent all topologically distinct types of complete intersection Calabi-Yau four-folds is provided and can be downloaded \href{http://www-thphys.physics.ox.ac.uk/projects/CalabiYau/Cicy4folds/index.html}{{\tt here}}. The manifolds have non-negative Euler characteristics in the range $0\leq \chi \leq 2610$. This data set will be of use in a wide range of physical and mathematical applications. Nearly all of these four-folds are elliptically fibered and are thus of interest for F-theory model building.}

\preprint{ITP--UH--04/13}
\arxivnumber{1303.1832}
\keywords{Differential and Algebraic Geometry, F-Theory, Superstring Vacua}

\begin{document}
\maketitle
\flushbottom

\section{Introduction}\seclabel{intro}

Calabi-Yau manifolds play an important role in several branches of mathematics and physics. Often one obstruction to progress in a given area is the lack of large data sets of example manifolds. In this paper, we take a step towards rectifying this situation by explicitly constructing and classifying a specific class of Calabi-Yau four-folds. This set consists of Calabi-Yau four-folds which can be realized as complete intersections in products of complex projective spaces (the CICYs), arguably the simplest construction of Calabi-Yau manifolds available. The data set we find consists of some 921,497 configuration matrices describing these Calabi-Yau four-folds and thus provides a large, explicit and easy to manipulate class of such manifolds. 

For Calabi-Yau three-folds, all possible distinct CICYs were classified in 1988 by Candelas et. al.~\cite{Candelas:1987kf}. By means of a computer algorithm, a list of 7890 configuration matrices was obtained. This data set has been immensely useful, particularly in the context of string theory, and is still used to this day. For example, more recently, freely-acting symmetries for CICY three-folds have been classified~\cite{Braun:2010vc} and a large class of heterotic string standard models has been constructed based on these manifolds~\cite{Anderson:2011ns,Anderson:2012yf}. The main purpose of the present paper is to carry out an analogous classification of CICY four-folds.

Calabi-Yau four-folds are of particular importance for the construction of four-di\-men\-sio\-nal ${\cal N}=1$ string vacua based on F-theory~\cite{Vafa:1996xn,Donagi:2008ca,Beasley:2008dc,Beasley:2008kw}. If the success of heterotic model building, where the systematic analysis of large classes of vacua has led to the discovery of many standard-like models~\cite{Anderson:2011ns}, is to be emulated in F-theory, large, accessible classes of Calabi-Yau four-folds will be required~\cite{Lynker:1998pb}. Moreover, for the application to F-theory, Calabi-Yau four-folds need to allow for an elliptic fibration structure, where the six-dimensional base manifold corresponds to the ``physical" space required in the compactification from ten to four dimensions and the torus fiber describes the variation of the axio-dilaton over this base space. As we will see, practically all of the CICY four-folds which arise from our classification allow for an elliptic fibration and are, therefore, of potential use for F-theory. 

In order to introduce some basic ideas and discuss elementary properties of CICY four-folds we would like to start with a prototypical example, given by the configuration matrix
\be\eqlabel{egconf}
 \left[\begin{array}{c|cc}1&1&1\\2&1&2\\3&0&4\end{array}\right] \; .
\ee
The notation is to be understood as follows. The first column of the matrix denotes the dimensions of the projective spaces whose product forms the ambient space into which the CICY is embedded. Here, this ambient space is $\CP^1 \times\CP^2\times \CP^3$. Each of the remaining columns denotes the multi-degree of a polynomial in the ambient projective coordinates. For the present example, we have two polynomials with multi-degrees $(1,1,0)$ and $(1,2,4)$, where the three entries refer to the degrees in the coordinates of $\CP^1$, $\CP^2$ and $\CP^3$, respectively. The CICY defined is the common zero locus of these polynomials. If we denote the $\CP^1$ coordinates by $x^i$, where $i=0,1$, the $\CP^2$ coordinates by $y^a$, where $a=0,1,2$ and the $\CP^3$ coordinates by $z^\alpha$, where $\alpha=0,\ldots ,3$, then these polynomials can be written as
\be
 p_1= \sum_{i,a}c_{ia}x^iy^a\; ,\qquad p_2=\sum_{i,\ldots,\delta}d_{iab\alpha\beta\gamma\delta}x^iy^ay^bz^\alpha z^\beta z^\gamma z^\delta\; ,
\ee
where $c_{ia}$ and $d_{iab\alpha\beta\gamma\delta}$ are complex coefficients. Hence, the configuration matrix~\eqref*{egconf} describes a family of CICYs parametrized by the space of coefficients in these polynomials. Fortunately, many of the basic properties, such as the Euler characteristic, do not depend on the specific choice of these coefficients but only on the configuration matrix. This feature is of course one of the strengths of the configuration notation and one of the main motivations for its introduction.

For the purpose of applications to F-theory, how do we identify the existence of an elliptic fibration structure for such a CICY four-fold?
In fact, the configuration matrix~\eqref*{egconf} represents an example of a CICY with an ``obvious" elliptic fibration, that is, a fibration which is consistent with the projective ambient space embedding. To see this we note that the first two rows of the configuration matrix~\eqref*{egconf} are given by
\be
 \left[\begin{array}{c|cc}1&1&1\\2&1&2\end{array}\right] 
\ee
and represent a Calabi-Yau one-fold, that is, a torus $T^2$. The full configuration~\eqref*{egconf} describes a CICY where this torus is fibered over the base space $\CP^3$. It turns out that this fibration has section. As we will show, all but 477 of our 921,497 CICY configuration matrices have an elliptic fibration of this kind, consistent with the projective embedding. Indeed, many of these have a large number of different such fibrations, many of them with sections. This means the number of physical F-theory compactifications which can be obtained from this data set is, in fact, much larger than 921,497.

\vspace{0.1cm}

Our approach for classifying CICY four-folds will broadly follow the algorithm for the classification of CICY three-folds set out in ref.~\cite{Candelas:1987kf}. However, the large scope of the project, reflected in the total number of configuration matrices and their maximal size, means that numerous efficiency improvements had to be made in order to complete the task in a reasonable amount of computing time. Moreover, some of the methods do not generalize from three- to four-folds and had to be modified appropriately. As an example, we mention the operation on configuration matrices referred to as ``splitting". It involves increasing the size of the configuration by breaking up a column of the original matrix into several summands and adding a $\CP^n$ factor to the ambient space. A crucial step in the classification algorithm is to decide whether or not a splitting is effective, that is, whether it leads to a topologically different manifold. Unfortunately, the effectiveness criterion for CICY three-folds developed in ref.~\cite{Candelas:1987kf} does not generalize to four-folds and a new criterion had to be found. 
The details of the classification algorithm, including an effectiveness criterion for four-fold splittings, and the main results of the classification will be described in the remainder of this paper. In a longer, companion paper to this article~\cite{paper2}, we will provide additional properties of the manifolds in this data set. This will include information on Hodge numbers, Chern classes, and the structure of elliptic fibrations and sections.

The paper is organized as follows. In the next section, we define the data set we will be studying in more detail and explain why a finite number of configuration matrices suffices to represent all CICY four-folds. Essentially, different configuration matrices can describe the same Calabi-Yau manifold, and all CICY four-folds are accounted for by a finite subset of the infinite number of possible configuration matrices. We obtain upper bounds on the size of the matrices that need be considered and provide a table of all possible ambient spaces that can occur in this finite list. To classify the different manifolds it is useful to compute the Euler characteristic $\chi$, which only depends on the configuration matrix. The formula for $\chi$ together with expressions for the Chern classes are introduced in \secref{chern_and_euler}. In \secref{remove_redundancies} different types of possible equivalences, which have been taken into account in the compilation of our list, are discussed. It is explained how they generalize known results for three-folds to four-folds and how they can be dealt with efficiently. In \secref{algorithm}, we describe in detail the algorithm that was used to compile our list. The results of running this algorithm are presented in \secref{results}. We provide a histogram of the different values for the Euler characteristic that occur in the list, discuss the question of how many topologically distinct manifolds are present and how many manifolds have an obvious fibration structure. We conclude in \secref{outlook}.

\section{Definitions and finiteness of the class}\seclabel{finiteness}

We begin with a general description of the CICY four-folds classified in this paper. Our notation and conventions largely follow the original papers on CICY three-folds~\cite{Hubsch:1986ny,Green:1986ck,Candelas:1987kf,Candelas:1987du} and ref.~\cite{Hubsch:1992nu}. We consider the complete intersection of $K$ polynomials $p_\alpha$ in a product of $m$ projective spaces $\CP^{n_1} \times\cdots\times \CP^{n_m}$ of total dimension $K+4=\sum_{r=1}^mn_r$. In the following, we use indices $r,s,\ldots =1,\ldots,m$ to label the projective ambient space factors $\CP^{n_r}$ and indices $\alpha,\beta,\dots =1,\ldots ,K$ to label the polynomials $p_\alpha$. Such manifolds are described by a \emph{configuration matrix}
\be\eqlabel{conf2}
 [{\bf n}|{\bf q}] \equiv \left[\begin{array}{c|ccc}n_1 & q^1_1&\dots&q^1_K\\ \vdots & \vdots&\ddots&\vdots\\ n_m & q^m_1&\hdots&q^m_K\\\end{array}\right] ,
\ee
with non-negative integer entries $q_\alpha^r$. The columns ${\bf q}_\alpha =(q_\alpha^r)_{r=1,\ldots ,m}$ of this matrix denote the multi-degrees of the defining polynomials $p_\alpha$. More precisely, the polynomial $p_\alpha$ is of degree $q_\alpha^r$ in $x_{r,i}$, the homogeneous coordinates of $\CP^{n_r}$. In order to ensure that this prescription defines a four-dimensional manifold, we demand that the $K$-form
\be
 \dd p_1 \wedge \cdots \wedge \dd p_K
\ee
is nowhere vanishing.

The configuration $[{\bf n}|{\bf q}]$ describes a family of CICYs redundantly parametrized by the space of coefficients in the polynomials $p_\alpha$. The strength of this notation rests on the fact that key properties of the manifolds defined in this way only depend on the configuration matrix and not on the specific choice of polynomial coefficients. Moreover, it was shown in ref.~\cite{Green:1986ck} that for every configuration a generic choice of coefficients defines a complete intersection manifold. In the following, we will not distinguish between the family $[{\bf n}|{\bf q}]$ and a specific member thereof.

In order for a configuration matrix~\eqref*{conf2} to define Calabi-Yau manifolds we must ensure the vanishing of the first Chern class which is equivalent to the conditions
\be\eqlabel{c1zero}
 \sum_{\alpha = 1}^K q_\alpha^r = n_r + 1
\ee
on each row of the configuration matrix.

The conditions on CICY configuration matrices stated so far are not particularly stringent and it is clear that the set of such matrices is infinite. However, different configuration matrices can describe the {\it same} Calabi-Yau four-fold. In order to arrive at a finite list classifying all topological types of CICY four-folds, we need to identify suitable equivalence relations between configurations and only keep one representative per class. 

The simplest example of such an equivalence relation stems from the following observation. The ordering of ambient space factors and polynomials in the configuration matrix is completely arbitrary. Therefore, two configuration matrices that differ only by permutations of rows or columns describe the same family of CICY four-folds. To reduce the occurrence of such permutations we will, in our algorithm, impose a \emph{lexicographic order} (with the entries $q_\alpha^r=0,1, 2,\ldots$ ordered by value) on the rows and columns~\cite{Candelas:1987kf}. It then suffices to consider only permutations of rows where the corresponding ambient space factors are the same.

Another relevant observation is that a polynomial linear in the coordinates of a single $\CP^n$ defines a sub-manifold $\CP^{n-1} \subset \CP^n$. This means that a multi-degree ${\bf q}_\alpha$ with a single non-zero entry $q_\alpha^r=1$ can be removed from a configuration matrix while simultaneously reducing the dimension $n_r$ to $n_r-1$. To exclude such cases, we will require the degree of a polynomial to be at least two if it depends on one projective space only. This is equivalent to the condition
\be\eqlabel{no_lin_poly}
 \sum_{r=1}^m q_\alpha^r \geq 2 \; , \qquad\qquad \forall \alpha = 1,\ldots,K \; ,
\ee
which we impose on all configuration matrices.

Further, we note that we are not interested in block-diagonal configuration matrices of the form
\be\eqlabel{prod_mfld}
 \left[\begin{array}{c|cc}1 & 2 & 0 \\ {\bf n} & 0 & {\bf q}\end{array}\right] \; .
\ee
The sub-configuration $[1|2]$ describes two points in $\CP^1$ and the above configuration is, therefore, equivalent to two copies of $[{\bf n}|{\bf q}]$. 

Now focus on configuration matrices with a fixed size, $(m,K)$. All such matrices can be generated by a two-step procedure that is well-suited for machine computation~\cite{Hubsch:1992nu}. First, one lists all $m$--dimensional integer vectors ${\bf n}$ with $n_r>0$, ordered such that $n_r \geq n_s$ if $r>s$ , which satisfy the dimensional constraint $\sum_{r=1}^m n_r = K + 4$. Second, for each ${\bf n}$, one lists all matrices ${\bf q}$ which satisfy \eqref{c1zero,no_lin_poly}, excluding matrices of the form~\eqref*{prod_mfld}. This is most easily done by starting from an initial configuration and shifting row-wise according to
\be
 [\ldots, q_\alpha^r, q_{\alpha+1}^r, \ldots] \quad\to\quad [\ldots, (q_\alpha^r + 1), (q_{\alpha+1}^r-1), \ldots] \; ,
\ee
while preserving the lexicographic order of rows and columns.

For a given dimension vector ${\bf n}$, this procedure clearly terminates. However, it is not clear that the complete algorithm will also terminate and lead to a finite list, since the list of vectors ${\bf n}$ is, a priori, unbounded. However, it has been observed~\cite{Green:1986ck} that beyond a certain upper limit in ${\bf n}$, every configuration matrix is equivalent, by the above relations, to a smaller matrix and, hence, does not need to be included. In this sense, only the \emph{minimal} configuration of a given manifold is kept in the list. More precisely, generalizing the arguments in ref.~\cite{Green:1986ck}, it can be shown that minimal CICY $d$--folds satisfy the bounds
\be\eqlabel{nfold_matrix_size_upper_bounds}
 p \leq \alpha \leq 2d \; , \qquad s \leq 3d \; .
\ee
Here, $s$ is the number of ambient $\CP^1$ factors and $p$ the number of ambient $\CP^n$ factors, with $n>1$. The quantity $\alpha$ is defined as $\alpha := \sum_{\{r\,|\, n_r>1\}} (n_r - 1)$, where the sum is over all ambient $\CP^n$ factors with $n>1$. Since this bounds the total number, $m$, of ambient projective spaces as well as the total ambient space dimension from above, the set of minimal configurations is finite. For CICY four-folds, we must set $d=4$ and hence the bounds become
\be\eqlabel{4fold_matrix_size_upper_bounds}
 p \leq \alpha \leq 8 \; , \qquad s \leq 12 \; .
\ee
There are 660 different possible ambient spaces that satisfy these bounds and they are presented in \tabref{tab:ambsp4}.
\begin{table}[t]
\begin{center}
\begin{tabular*}{0.75\textwidth}{|@{\extracolsep{\fill}}l|c|c|c|r|}
\hline
\emph{Space}													&	$g$								& $f_{\text{max}}$ 	& $N_{\rm ex}$ & \emph{Number}\\\hline
$(\CP^1)^f \CP^9$											&										&		$5$							&	$0$					&		$6$		\\
$(\CP^1)^f (\CP^5)^2$									&										&		$6$							&	$0$					&		$7$		\\
$(\CP^1)^f \CP^4 \CP^6$								&										&		$6$							&	$0$					&		$7$		\\
$(\CP^1)^f \CP^3 \CP^7$								&										&		$6$							&	$0$					&		$7$		\\
$(\CP^1)^f (\CP^2)^g \CP^8$						&	$0\rightarrow 1$	&		$6$							&	$1-g$				&		$14$	\\
$(\CP^1)^f \CP^3 (\CP^4)^2$						&										&		$7$							&	$0$					&		$8$		\\
$(\CP^1)^f (\CP^3)^2 \CP^5$						&										&		$7$							&	$0$					&		$8$		\\
$(\CP^1)^f (\CP^2)^g \CP^4 \CP^5$			&	$0\rightarrow 1$	&		$7$							&	$1-g$				&		$16$	\\
$(\CP^1)^f (\CP^2)^g \CP^3 \CP^6$			&	$0\rightarrow 1$	&		$7$							&	$1-g$				&		$16$	\\
$(\CP^1)^f (\CP^2)^g \CP^7$						&	$0\rightarrow 2$	&		$7$							&	$2-g$				&		$24$	\\
$(\CP^1)^f (\CP^3)^4$									&										&		$8$							&	$0$					&		$9$		\\
$(\CP^1)^f (\CP^2)^g (\CP^3)^2 \CP^4$	&	$0\rightarrow 1$	&		$8$							&	$1-g$				&		$18$	\\
$(\CP^1)^f (\CP^2)^g (\CP^4)^2$				&	$0\rightarrow 2$	&		$8$							&	$2-g$				&		$27$	\\
$(\CP^1)^f (\CP^2)^g \CP^3 \CP^5$			&	$0\rightarrow 2$	&		$8$							&	$2-g$				&		$27$	\\
$(\CP^1)^f (\CP^2)^g \CP^6$						&	$0\rightarrow 3$	&		$8$							&	$3-g$				&		$36$	\\
$(\CP^1)^f (\CP^2)^g (\CP^3)^3$				&	$0\rightarrow 2$	&		$9$							&	$2-g$				&		$30$	\\
$(\CP^1)^f (\CP^2)^g \CP^3 \CP^4$			&	$0\rightarrow 3$	&		$9$							&	$3-g$				&		$40$	\\
$(\CP^1)^f (\CP^2)^g \CP^5$						&	$0\rightarrow 4$	&		$9$							&	$4-g$				&		$50$	\\
$(\CP^1)^f (\CP^2)^g (\CP^3)^2$				&	$0\rightarrow 4$	&		$10$						&	$4-g$				&		$55$	\\
$(\CP^1)^f (\CP^2)^g \CP^4$						&	$0\rightarrow 5$	&		$10$						&	$5-g$				&		$65$	\\
$(\CP^1)^f (\CP^2)^g \CP^3$						&	$0\rightarrow 6$	&		$11$						&	$6-g$				&		$82$	\\
$(\CP^1)^f (\CP^2)^g$									&	$0\rightarrow 8$	&		$12$						&	$8-g$				&		$108$	\\\hline
\end{tabular*}
\caption{All possible ambient spaces for CICY four-folds are shown in this table. These $660$ ambient manifolds fall into classes according to the number of $\CP^1$- and $\CP^2$-factors. The third column gives the excess number $N_{\rm ex} = \sum_{r=1}^{m} (n_r + 1) - 2K$. It vanishes when all the columns sum to two which, from \eqref{no_lin_poly}, is the minimal non-trivial value. A large value of $N_{\rm ex}$ generally means that there are many ways to construct inequivalent configuration matrices for a given ambient space. The minimum number of $\CP^1$ factors is zero except for 
$(\CP^1)^f$ where $f_{\text{min}} = 5$,
$(\CP^1)^f \CP^2$ where $f_{\text{min}} = 3$,
$(\CP^1)^f \CP^3$ where $f_{\text{min}} = 2$,
$(\CP^1)^f \CP^4$ where $f_{\text{min}} = 1$ and
$(\CP^1)^f (\CP^2)^2$ where $f_{\text{min}} = 1$. This table follows the format used in ref.~\cite{Candelas:1987kf}.}
\tablabel{tab:ambsp4}
\end{center}
\end{table}

As will be explained in \secref{remove_redundancies}, it is possible to employ further techniques, beyond those discussed here to remove redundant descriptions of CICYs. This will lead to the refined, more efficient algorithm described in \secref{algorithm}. However, as we will see, the simple method outlined in this section still serves a useful purpose as the first, initiating step of the full algorithm.

\section{Chern classes and Euler characteristic}\seclabel{chern_and_euler}

To implement more advanced methods for redundancy removal, we require explicit expressions for some of the topological properties of complete intersection manifolds. For this reason, we review the explicit formulae for the Euler characteristic, which is of particular importance, and the Chern classes. These formulae will be presented for general complete intersection manifolds with configuration matrix $[{\bf n}|{\bf q}]$ which do not necessarily have to satisfy the Calabi-Yau condition~\eqref*{c1zero}.

We begin with the total Chern class which is given by the expression~\cite{Green:1986ck}
\be\eqlabel{totalChernclass}
 c( [{\bf n}|{\bf q}] ) = \frac{\prod_{r=1}^m (1 + J_r)^{n_r+1}}{\prod_{\alpha=1}^K (1 + \sum_{s=1}^m q_\alpha^s J_s)} \; ,
\ee
where $J_r$ denotes the K\"ahler form of the $r$-th ambient projective space $\CP^{n_r}$, normalized in the standard way such that
\be\eqlabel{Pnorm}
	\int_{\CP^{n_r}} J_r^{n_r} = 1 \; .
\ee
Expanding \eqref{totalChernclass} yields explicit formulae for the first four Chern classes. They are given by
\begin{align}
	c_1([\mathbf{n}|\mathbf{q}])&=c_1^r J_r = \left[n_r+1-\sum_{\alpha=1}^Kq^r_\alpha\right]J_r \; , \eqlabel{c1} \\
	c_2([\mathbf{n}|\mathbf{q}])&= c_2^{rs} J_r J_s = 
		\oneon{2} \left[-(n_r+1)\delta^{rs} + \sum_{\al=1}^K q_\al^r q_\al^s + c_1^r c_1^s \right] J_r J_s \; , \eqlabel{c2} \\
	c_3([\mathbf{n}|\mathbf{q}])&= c_3^{rst} J_r J_s J_t = 
		\oneon{3} \left[(n_r+1)\delta^{rst} - \sum_{\al=1}^K q_\al^r q_\al^s q_\al^t + 3 c_1^r c_2^{st} - c_1^r c_1^s c_1^t \right] J_r J_s J_t \; , \eqlabel{c3} \\
	c_4 ([\mathbf{n}|\mathbf{q}]) &= c_4^{rstu} J_r J_s J_t J_u = \oneon{4}
		\left[ -(n_r+1)\delta^{rstu} + \sum_{\al=1}^K q_\al^r q_\al^s q_\al^t q_\al^u + 2 c_2^{rs} c_2^{tu} \right. \nonumber 
		\\ & \left. \qquad\qquad\qquad\qquad\qquad\qquad\qquad + 4 c_1^r c_3^{stu} - 4 c_1^r c_1^s c_2^{tu} + c_1^r c_1^s c_1^t c_1^u \vphantom{\sum_{\al=1}^K} \right] J_r J_s J_t J_u \; . \eqlabel{c4}
\end{align}
Here, the multi-index Kronecker delta is defined to be $\delta^{r_1 \ldots r_n} = 1$ if $r_1 = r_2 = \ldots = r_n$ and zero otherwise. For a configuration to describe a family of Calabi-Yau manifolds we need $c_1([{\bf n}|{\bf q}])=0$ which leads to the Calabi-Yau constraint~\eqref*{c1zero} presented earlier. In this case, the above equations for the higher Chern classes simplify substantially since all terms proportional to the first Chern class can be dropped. 

The fourth Chern class is related to the Euler characteristic $\chi$ by a variant of the Gauss-Bonnet formula
\be\eqlabel{Euler_c4}
	\chi ([{\bf n}|{\bf q}]) = \int_{[{\bf n}|{\bf q}]} c_4([{\bf n}|{\bf q}]) \; .
\ee
An integration of a top-form $\omega$ over $[{\bf n}|{\bf q}]$ is evaluated by pulling it back to an integration over the ambient space ${\cal A} = \CP_1^{n_1} \times\cdots\times \CP_m^{n_m}$ using
\be\eqlabel{mudef}
	\int_{[\mathbf{n}|\mathbf{q}]} \omega = \int_{\cal A} \omega \wedge\mu_{[{\bf n}|{\bf q}]} \; ,\qquad 
	\mu_{[{\bf n}|{\bf q}]} \equiv \bigwedge_{\al=1}^K\left(\sum_{r=1}^mq^r_\al J_r\right) ,
\ee
and the normalizations~\eqref*{Pnorm} of the K\"ahler forms $J_r$. The $(K,K)$-form $\mu_{[{\bf n}|{\bf q}]}$ is the Poincar\'e dual to the sub-manifold $[{\bf n}|{\bf q}]$ in the ambient space ${\cal A}$.

The explicit formula for the Euler characteristic $\chi$ of a four-fold configuration $[{\bf n}|{\bf q}]$ is then given by
\be
 \chi ([{\bf n}|{\bf q}]) = \left[c_4([{\bf n}|{\bf q}])\wedge \mu_{[{\bf n}|{\bf q}]}\right]_{\rm top} \eqlabel{chi}
 \ee
where the subscript ``top'' means that the coefficient of the volume form $J_1^{n_1} \wedge\cdots\wedge J_m^{n_m}$ of ${\cal A}$ should be extracted from the enclosed expression. 

For Calabi-Yau manifolds, vanishing of the first Chern class, $c_1^r = 0$, implies that $(n_r + 1) \leq \sum_{\alpha=1}^K (q_\alpha^r)^\ell$, for $\ell = 1,2,3,\ldots$, and hence $c_2^{rs} \geq 0$, $c_4^{rstu} \geq 0$. This shows that $\chi ([{\bf n}|{\bf q}]) \geq 0$ for all CICY four-folds.

\section{Equivalent configurations and redundancy removal}\seclabel{remove_redundancies}

After this preparation, we can now discuss more refined equivalence relations between configuration matrices. It will then be a simple matter, in the next section, to construct an improvement on the ``naive algorithm'' given in \secref{finiteness}. There are several different ways in which two configuration matrices can be equivalent:
\paragraph{I. Permutations of rows and columns.}
As we have already discussed, two configuration matrices are equivalent if they differ only by a permutation of rows or columns. The resulting redundancy is partially resolved by imposing the aforementioned lexicographic order on the rows and columns~\cite{Candelas:1987kf}. However, a residual redundancy remains. A ``brute force'' procedure to remove this redundancy is to generate all row and column permutations of a matrix and compare with the candidate equivalent configuration. For the larger CICY configuration matrices which appear in our classification, this eventually gets out of hand, due to the exponential growth of the number of permutations with matrix size.

An alternative method which is more efficient, particularly for large matrix size, works as follows. Consider two configurations, $[{\bf n}|{\bf q}]$ and $[{\bf n}|{\bf \tilde{q}}]$, of the same size. First we impose a sequence of necessary conditions for equivalence in order to identify inequivalent configurations efficiently. The algorithm is stopped as soon as non-equivalence is established. The first necessary condition is that the tallies of numbers in each row and column should coincide for two matrices related by row or column permutations. Hence, if the tally disagrees the matrices are inequivalent. In the second step, we compare the trace and eigenvalues of the $m\times m$ square matrices ${\bf M}={\bf q} {\bf q}^T$ and ${\bf \tilde{M}}={\bf \tilde{q}} {\bf \tilde{q}}^T$. If either disagrees the matrices are inequivalent.

For configurations which pass these tests we have to find a necessary and sufficient criterion for equivalence. To this end consider $O(m)$ matrices ${\bf R}$ and ${\bf \tilde{R}}$ diagonalizing ${\bf M}$ and ${\bf \tilde{M}}$, that is, ${\bf R}^T {\bf M} {\bf R} = {\bf \tilde{R}}^T {\bf \tilde{M}} {\bf \tilde{R}} = {\rm diag}(a_1,\ldots,a_m)$. In addition, we assume that the eigenvalue spectrum $\{a_r\}$ is non-degenerate.\footnote{If the spectrum happens to be degenerate we can either modify the configuration matrices ${\bf q}$ and $\tilde{\bf q}$ in a way that does not affect equivalence but may change the spectrum, for example by adding the same constant to each entry, or use the brute force method described earlier.} The crucial observation is then that, given a fixed order of the eigenvalues, the matrices ${\bf R}$ and ${\bf \tilde{R}}$ are essentially unique apart from a sign choice for each eigenvector. This sign ambiguity can be fixed by demanding that 
\be
 \sum_{r=1}^m {\bf R}_{rs} > 0 \; , \qquad \sum_{r=1}^m {\bf \tilde{R}}_{rs} > 0\;,\qquad \forall s=1,\ldots,m \; .
\ee
Given these sign conventions we then compute the matrix ${\bf P} = {\bf \tilde{R}} {\bf R}^T$ and check if it is a permutation matrix. If it is not, the configurations are inequivalent. If it is, we compute ${\bf q'} = {\bf P}^T {\bf \tilde{q}}$ and check if it has the same column vector set as ${\bf q}$. If it does, the two configurations are equivalent, otherwise they are not.

All of the above can be efficiently implemented in Mathematica. The full proof that this procedure is indeed necessary and sufficient for deciding the equivalence of two configurations will be given in the forthcoming longer publication~\cite{paper2}.
\paragraph{II. Ineffective splittings.}
The \emph{splitting principle}~\cite{Candelas:1987kf} provides an efficient method of generating new configurations from old ones. It plays a key role in the algorithm to generate the full list of CICY configurations, as will be explained in \secref{algorithm}. As we shall see in what follows, deciding whether or not a four-fold splitting is effective, that is, whether it leads to a new manifold, cannot be accomplished by a simple generalization of the three-fold criterion and requires some new ideas.

A general $\CP^n$ splitting is defined as a relation of the form
\be\eqlabel{Pnsplit}
 \left[\begin{array}{c|cc}{\bf n} & \displaystyle\sum_{a=1}^{n+1} {\bf u}_a & {\bf q}\end{array}\right] \longleftrightarrow\; \left[\begin{array}{c|ccccc}n & 1 & 1 & \cdots & 1 & 0\\{\bf n} & {\bf u}_1 & {\bf u}_2 & \cdots & {\bf u}_{n+1} & {\bf q}\end{array}\right] \; .
\ee
Read from left to right this correspondence is termed \emph{splitting} while its inverse is called \emph{contraction}. When the two configurations describe the same underlying manifold, the splitting is called \emph{ineffective}, otherwise it is referred to as an \emph{effective} splitting.

To decide whether or not the two configurations in~\eqref*{Pnsplit} describe the same underlying manifold, we first note that these two manifolds share common loci in their complex structure moduli space, the so called determinantal variety. To see this, introduce homogeneous coordinates ${\bf x}=(x_i)_{i=0,\ldots ,n}$ for the additional $\CP^n$ which arises in the splitting and a matrix ${\bf F} = (f_{ai})$ of polynomials $f_{ai}$ with multi-degrees ${\bf u}_a$. Then, the zero locus of the first $n+1$ polynomials in the split configuration in~\eqref*{Pnsplit} can be written as ${\bf F}{\bf x}=0$. Evidently, this equation has a solution in $\CP^n$ if and only if $p\equiv {\rm det}({\bf F})=0$. The polynomial $p$ has multi-degree ${\bf u}=\sum_{a=1}^{n+1}{\bf u}_a$ and is a specific instance of the first defining polynomial of the contracted configuration in~\eqref*{Pnsplit}. Together with the polynomials specified by ${\bf q}$ it defines the determinantal variety. The question then becomes whether or not this determinantal variety is smooth. If it is, the two configurations can be smoothly deformed into each other and, hence, represent the same topological type of Calabi-Yau manifolds. In this case, the splitting is ineffective. Otherwise, that is, when the determinantal variety has a non-trivial singular locus, they describe different manifolds and the splitting is effective.

For CICY three-fold splittings, the singular locus of the determinantal variety is a zero-dimensional space. That is, it can either be the empty set or a collection of points. It turns out that the number of singular points is counted, up to a non-zero numerical factor, by the difference of Euler characteristics between the original and the split configuration. This leads to the simple rule that two three-fold configurations, related by splitting as in~\eqref*{Pnsplit}, are equivalent if and only if they have the same Euler characteristic~\cite{Candelas:1987kf}.

For a CICY four-fold, the singular locus of the determinantal variety has a more complicated structure. As was first noted in ref.~\cite{Brunner:1996bu}, four-fold splittings have a different local degeneration structure than three-fold splittings. The determinantal variety of a CICY four-fold splitting becomes singular on a complex curve. The Euler characteristic of this curve is still proportional, with a non-zero factor, to the difference of Euler characteristics between the two configurations involved. This means that a four-fold splitting which changes the Euler characteristic is definitely effective. If the splitting preserves the Euler characteristic, however, then we only know that the singular locus must have vanishing Euler characteristic. This means that the singular locus could either be the empty set or a collection of tori. In the case of CICY four-folds, therefore, it is possible to have effective splittings at constant Euler characteristic. Clearly, to detect such effective splittings which preserve the Euler characteristic we need additional criteria.

For $\CP^1$ splittings between CICY four-folds, a necessary and sufficient criterion can be obtained as follows. In this case, the one-dimensional singular locus of the determinantal variety can be described as a complete intersection, associated to the configuration matrix $S \equiv \left[\begin{array}{c|ccccc}{\bf n} & {\bf u_1} & {\bf u_1} & {\bf u_2} & {\bf u_2} & {\bf q}\end{array}\right]$. We denote by $\mu_S$ the form Poincar\'e-dual to this singular locus in the ambient space ${\cal A}$, defined analogously to \eqref{mudef}, and by $J$ a K\"ahler form on ${\cal A}$. A convenient choice for this K\"ahler form is $J=\sum_{r=1}^mJ_r$. Then, the volume of the singular locus can be calculated by
\be\eqlabel{VolX}
 \mathrm{Vol}(S) = \int_S J = \int_{\cal A} J \wedge \mu_S =\left[J\wedge \mu_S\right]_{\rm top} \; ,
\ee
where the subscript ``top" refers to the coefficient of the top form $J_1^{n_1} \wedge\cdots\wedge J_m^{n_m}$ of ${\cal A}$, as before.
With the expressions for $J$ and $\mu_S$ readily available, this allows for an explicit calculation of the volume, using the normalizations~\eqref*{Pnorm}. Clearly, the singular set $S$ is empty and, hence, the splitting ineffective, if and only if this volume vanishes. There is a trivial but helpful re-formulation of this criterion in terms of the associated zero-dimensional configuration $S' \equiv \left[\begin{array}{c|cccccc}{\bf n} & {\bf u_1} & {\bf u_1} & {\bf u_2} & {\bf u_2} & {\bf q} & {\bf 1}\end{array}\right]$, where ${\bf 1}$ denotes a column with all entries $1$. Then, for the choice of K\"ahler form $J=\sum_{r=1}^mJ_r$ it follows that
\be
 \chi(S') = \int_{S'} c_0 = \int_{\cal A} \mu_{S'} = \int_{\cal A} \mu \wedge \left( \sum_{r=1}^m J_r \right)={\rm Vol}(S) \; .
\ee
Hence, the splitting is effective if and only if $\chi(S')\neq 0$. 

Unfortunately, for higher $\CP^n$ splittings, $n>1$, the singular locus cannot be described as a complete intersection. Hence, the above method cannot be applied and we have to rely on a different approach. As before, the first step is to compute the change of the Euler characteristic using \eqref{chi}. If the Euler characteristic changes, we have an effective splitting. Otherwise, we consider the following splittings between \emph{non-Calabi-Yau} three-folds
\be\eqlabel{Pnsplit_assoc_divisorsplit}
 \left[\begin{array}{c|ccc}{\bf n} & \displaystyle\sum_{a=1}^{n+1} {\bf u}_a & {\bf q} & {\bf e}_i\end{array}\right] \;\longleftrightarrow\;
 \left[\begin{array}{c|cccccc}n & 1 & 1 & \cdots & 1 & 0 & 0 \\ {\bf n} & {\bf u}_1 & {\bf u}_2 & \cdots & {\bf u}_{n+1} & {\bf q} & {\bf e}_i\end{array}\right] \; .
\ee
They are related to the original four-fold splitting~\eqref*{Pnsplit} by adding one additional column, given by a standard $m$--dimensional unit vector ${\bf e}_i$, to both configuration matrices. The singular locus of these three-fold splittings consists of points whose number is proportional to the change in Euler characteristic.
With the equations provided in \secref{chern_and_euler}, we find that the change of Euler characteristic for each ${\bf e}_i$ is given by
\be
 \Delta\chi_i = 2 \Bigg[ \Big\{ \sum_{a<b} \hat{u}_a^2 \hat{u}_b^2 + \mathop{\mathop{\sum_{a\neq b}}_{a\neq c}}_{b<c} \hat{u}_a^2 \hat{u}_b \hat{u}_c + 2 \sum_{a<b<c<d} \hat{u}_a \hat{u}_b \hat{u}_c \hat{u}_d \Big\} \wedge J_i \wedge \mu_{[{\bf n}|{\bf q}]} \Bigg]_{\rm top} \; ,
\ee
where $\hat{u}_a := \sum_{r=1}^m u_a^r J_r$.
Of course, the singular points associated to the three-fold splittings~\eqref*{Pnsplit_assoc_divisorsplit} are precisely the intersections of the four-fold singular locus (a complex curve) with the hyperplanes defined by the additional ${\bf e}_i$ column. Hence, if the Euler characteristic changes for at least one ${\bf e}_i$ the four-fold singular locus must be non-empty and the splitting is effective. Conversely, if the difference of Euler characteristics vanishes for all ${\bf e}_i$, that is, none of the hyperplanes intersects the four-fold singular locus, then this locus must be empty and the splitting is ineffective.

In general, if two configurations are found to be related by an ineffective splitting, they describe the same underlying manifold and only the contracted matrix (that is, the matrix on the left hand side of~\eqref*{Pnsplit}) will be kept in our list.
\paragraph{III. Identities.}
Numerous identities between sub-configurations of CICYs have been uncovered and discussed in ref.~\cite{Candelas:1987kf}. For a few of them, only heuristic arguments exist. In the compilation of our list, we have only used those identities that have been proved rigorously and that commute with splitting, namely:
\begin{center}
\begin{longtable}{ l l l }\hline\endfirsthead\\
 (II) (i) & $[2|2] = \CP^1$ & $\left[\begin{array}{c|cc}2 & 2 & {\bf a} \\ {\bf n} & 0 & {\bf q}\end{array}\right] = \left[\begin{array}{c|c}1 & 2{\bf a}\\ {\bf n} & {\bf q}\end{array}\right]$  \\[4ex]\hline\\
 (II) (ii) & $\left[\begin{array}{c|c}1 & 1 \\ 1 & 1\end{array}\right] = \CP^1$ & $\left[\begin{array}{c|cc}1 & 1 & {\bf a} \\ 1 & 1 & {\bf b}\\ {\bf n} & 0 & {\bf q}\end{array}\right] = \left[\begin{array}{c|c}1 & {\bf a}+{\bf b}\\ {\bf n} & {\bf q}\end{array}\right]$ \\[5ex]\hline\\
 (III) (i) & $[3|2] = \CP^1 \times\CP^1$ & $\left[\begin{array}{c|cc}3 & 2 & {\bf a} \\ {\bf n} & 0 & {\bf q}\end{array}\right] = \left[\begin{array}{c|c}1 & {\bf a}\\ 1 & {\bf a}\\ {\bf n} & {\bf q}\end{array}\right]$ \\[5ex]\hline\\
 (III) (ii) & $\left[\begin{array}{c|c}1 & 2 \\ 2 & 1\end{array}\right] = \CP^1 \times\CP^1$ & $\left[\begin{array}{c|cc}1 & 2 & 0 \\ 2 & 1 & {\bf a}\\ {\bf n} & 0 & {\bf q}\end{array}\right] = \left[\begin{array}{c|c}1 & {\bf a}\\ 1 & {\bf a}\\ {\bf n} & {\bf q}\end{array}\right]$ \\[5ex]\hline\\
 (III) (v) $\quad$ & $\left[\begin{array}{c|cc}2 & 2 & 1 \\ 2 & 1 & 1\end{array}\right] = \left[\begin{array}{c|c}1 & 2 \\ 2 & 2\end{array}\right]$ $\quad$ & $\left[\begin{array}{c|ccc}2 & 2 & 1 & 0 \\ 2 & 1 & 1 & {\bf a}\\ {\bf n} & 0 & 0 & {\bf q}\end{array}\right] = \left[\begin{array}{c|cc}1 & 2 & 0 \\ 2 & 2 & {\bf a}\\ {\bf n} & 0 & {\bf q}\end{array}\right]$ \\[5ex]\hline
\end{longtable}
\end{center}
The first column provides the labeling of the identities used in ref.~\cite{Candelas:1987kf}. The second and third columns state the basic identity and its application to the full configuration matrix, respectively. The identities are used from left to right, that is, whenever a matrix matches the pattern on the left hand side, it is replaced by the matrix on the right hand side. The proof of the basic identities in the second column is facilitated by the fact that these are either identities between one-folds or between two-folds of positive first Chern class. Both sets of manifolds are classified by their Euler characteristics, which can be computed straightforwardly by using the formulae of \secref{chern_and_euler}.

\vspace{0.1cm}

This concludes the list of equivalence relations we will be using in our classification algorithm. Their application greatly reduces the number of repetitions in our final list of CICY four-folds. However, they do not represent an exhaustive list of identities. It is to be expected that our list of CICY four-folds still contains some repetitions. This is indeed the case for the list of 7890 CICY three-folds and has been explicitly checked in ref.~\cite{Anderson:2008uw}, using Wall's theorem~\cite{Hubsch:1992nu}. For our CICY four-fold list the obvious course of action is to compute topological quantities in order to discriminate between inequivalent configurations and to determine a lower bound for the number of inequivalent four-fold CICYs. Useful topological quantities in question include the Euler characteristic, Chern classes, Hodge numbers and intersection numbers. In the present paper, we will only explicitly use the Euler characteristic for this purpose. A more complete discussion which includes the other quantities will be presented in the companion paper~\cite{paper2}. However, the experience with CICY three-folds suggests that the number of inequivalent configurations is of the same order of magnitude as the total number of configurations in the list.

\section{The algorithm}\seclabel{algorithm}

In \secref{finiteness}, we have described a simple and finite algorithm to directly generate all possible configuration matrices. It turns out that this naive algorithm is prohibitively slow and requires a computation time which is unfeasibly long. In this section, we use an adapted version of an algorithm first devised by Candelas et.~al.~\cite{Candelas:1987kf} for CICY three-folds. The basic idea is to employ the splitting principle in order to generate new CICY configuration matrices starting from a relatively small initial set. 

In the first step of the algorithm, we compile a list of all configuration matrices in ambient spaces that do not contain any $\CP^1$ factors. This is done using the naive algorithm of \secref{finiteness}. There are 62 such ambient spaces out of the 660 listed in \tabref{tab:ambsp4}. A new matrix is only added to the list if it is not related by row or column permutations to a matrix already contained in the list. After about 987 CPU hours,\footnote{All CPU hours stated in this paper refer to times measured on a Linux cluster at the ITP, Leibniz Universit\"at Hannover, consisting of contemporary desktop computers with CPUs ranging from Intel Core Duo 2 GHz to Intel Quad Core i5 3.1 GHz.} a list $L_0$ consisting of 9522 configuration matrices in ambient spaces without $\CP^1$ factors is produced. This list is then subjected to a routine we will refer to as the \emph{second filter}. This filter takes a list of matrices and removes the three different types of redundancies described in sections~\ref{sec:remove_redundancies}.I--III as well as matrices of the form~\eqref*{prod_mfld}. The second filter routine thus produces a minimal version (``minimal'' in the sense of both the number of matrices \emph{and} the size of each individual matrix) of the input list. When applied to $L_0$, it yields a reduced list $L'_0$ containing 4898 matrices.

Since the identities listed in \secref{remove_redundancies}.III have been applied, the list $L'_0$ does contain some matrices with $\mathbb{P}^1$ factors in their ambient spaces. In particular some matrices with rows of the form $\left[\begin{array}{c|cccc}1 & 2 & 0 & \cdots & 0\end{array}\right]$ are present. The only type of matrices missing from this list are those that contain one or more rows of the form $\left[\begin{array}{c|ccccc}1 & 1 & 1 & 0 & \cdots & 0\end{array}\right]$. According to the splitting relation~\eqref*{Pnsplit}, these matrices must be related to the matrices in $L'_0$ by contraction. Conversely, the full list can be produced by repeatedly performing $\CP^1$ splittings in all possible ways on the matrices in $L'_0$.

The first complete $\CP^1$ splitting of $L'_0$ yields a list $L_1$ consisting of 28823 matrices. The union of $L'_0$ and $L_1$ is then subjected to the second filter routine. The output is a list $L'_1$. It contains $L'_0$ plus 25222 new matrices making a total of 30120. Afterwards, the set difference $\Delta_1 = L'_1 \setminus L'_0$ is split in all possible ways to obtain a list $L_2$ and the union $L'_1 \cup L_2$ is subjected to the second filter routine to yield a list $L'_2$. This is repeated until no more new matrices are produced. The inequality~\eqref*{4fold_matrix_size_upper_bounds} guarantees that the algorithm terminates after $L'_{12}$ at the latest. In the actual execution of the algorithm, it turns out that already after $L'_{11}$, all splittings become ineffective. Hence, $L'_{11}$ represents the final result.

A logic flowchart depicting the steps of the algorithm is shown in \figref{algorithm_flowchart}.
\begin{figure}
 \begin{center}
  \begin{tikzpicture}[node distance=3.5cm, auto, >=stealth]
   \node (a)                                            {{\footnotesize start}};
   \node[block] (b)  [right of=a, node distance=2.6cm]  {{\footnotesize ``naive algorithm''}};
   \node[block] (c)  [right of=b, node distance=3.8cm]  {{\footnotesize 2ndfilter($L_0$)}};
   \node[block] (d)  [right of=c, node distance=3.8cm]  {{\footnotesize splitting($L'_0$)}};
   \node[block] (e)  [below of=b, node distance=2.5cm]  {{\footnotesize 2ndfilter($L'_0 \cup L_1$)}};
   \node[block] (f)  [right of=e, node distance=3.8cm]  {{\footnotesize splitting($\Delta_1$)}};
   \node[block] (g)  [right of=f, node distance=3.8cm]  {{\footnotesize 2ndfilter($L'_1 \cup L_2$)}};
   \node (h)  [right of=g, node distance=2.6cm]  {$\ldots$};
   
   \draw[->] (a) --                                      (b);
   \draw[->] (b) -- node[above] {{\footnotesize $L_0$}}  (c);
   \draw[->] (c) -- node[above] {{\footnotesize $L'_0$}} (d);
   \draw[->] (d.south) to [out=210,in=20] node[above] {{\footnotesize $L_1$}} (e.north);
   \draw[->] (e) -- node[above] {{\footnotesize $L'_1$}} (f);
   \draw[->] (f) -- node[above] {{\footnotesize $L_2$}}  (g);
   \draw[->] (g) -- node[above] {{\footnotesize $L'_2$}} (h);
  \end{tikzpicture}
  \caption{Logic flowchart of the algorithm described in \secref{algorithm}. The boxes label the routines executed at each step and the arguments in parentheses are the input for the routines. The ``naive algorithm'' is presented in \secref{finiteness}. The second filter routine is denoted ``2ndfilter'' for brevity. By ``splitting'', we refer to a routine which carries out all possible $\CP^1$ splittings on the matrices of the input list. The output lists are displayed above the arrows. The sets $\Delta_i$ are defined as $\Delta_i := L'_i \setminus L'_{i-1}$. The algorithm terminates after 11 consecutive splittings with the routine 2ndfilter$(L'_{10} \cup L_{11})$, which produces the final output $L'_{11}$.}
  \figlabel{algorithm_flowchart}
 \end{center}
\end{figure}
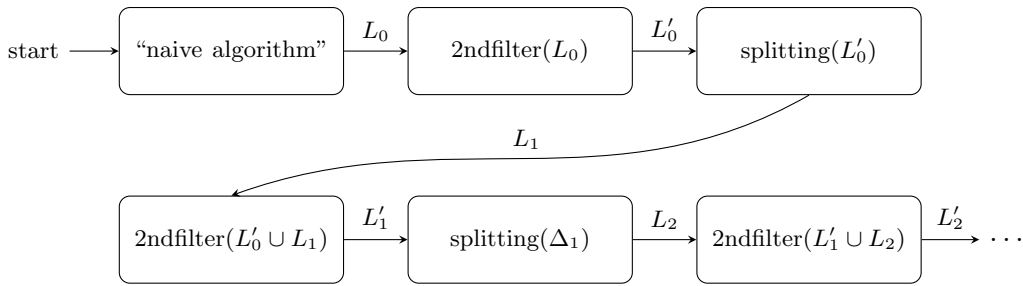
%

\section{Results}\seclabel{results}

Before we describe the results of our CICY four-fold classification, we first check that our implementation of the algorithm described in \secref{algorithm} successfully reproduces the known list of CICY three-folds. The original CICY three-fold list compiled in ref.~\cite{Candelas:1987kf} can be obtained from~\cite{cicylist}. It consists of 7890 CICY three-fold configuration matrices which include 22 direct product manifolds and 7868 spaces that cannot be written as direct products. A comparison with the list produced by our code shows a perfect match. The total CPU time to compile this list was just 72 minutes.

We now present our main result, a complete classification of CICY four-folds. The list contains 921,497 configuration matrices ranging up to a matrix size of $16\times 20$. The total required CPU time was 7487 hours, that is about 312 days on a single CPU.\footnote{In fact, we have used up to 20 CPUs in parallel for the splitting of matrices in order to shorten the running time.} A subset of 15813 matrices corresponds to product manifolds. These fall into four types as listed in the following table:
\begin{center}
\begin{tabular}{ l l l }\hline
 \emph{Type}\hspace{15mm} & \emph{Number of matrices} & \emph{Euler characteristic $\chi$}\\\hline
 $T^8$ & 5 & 0 \\
 $T^2 \times$CY$_3$ & 15736 & 0 \\
 $T^4 \times K3$ & 27 & 0 \\
 $K3 \times K3$ & 45 & 576 \\\hline
\end{tabular}
\end{center}
The Euler characteristic of these direct product manifolds follows from $\chi(M \times N) = \chi(M) \cdot \chi(N)$ together with $\chi(T^n) = 0$ and $\chi(K3) = 24$. The numbers of these different types of direct product matrices in the second column can be explained as follows. The algorithm produces two different configuration matrices for $T^2$, namely
\be\eqlabel{T2confs}
 [2|3] \qquad\text{and}\qquad \left[\begin{array}{c|c}1 & 2\\ 1 & 2\end{array}\right] \; .
\ee
For $K3$, 9 different configuration matrices are generated and the number of non block-diagonal CICY three-fold configurations is 7868. There are clearly five inequivalent ways to combine the two $T^2$ configurations~\eqref*{T2confs} into a $T^8$, the same as the dimension of the space of order four polynomials in two variables. The number of direct product matrices for $T^2 \times\text{CY}_3$ simply follows from $\#(T^2 \times\text{CY}_3) = \#(T^2) \cdot \#(\text{CY}_3) = 2 \cdot 7868 = 15736$. Similarly, $\#(T^4 \times K3) = \#(T^4) \cdot \#(K3) = 3 \cdot 9 = 27$. Finally, $\#(K3 \times K3) = \frac{9\cdot 10}{2} = 45$. Adding the numbers of the first three rows yields 15768. This precisely matches the number of matrices with Euler characteristic equal to zero and hence, all of them are product manifolds. The Euler characteristic 576 arises 2632 times in the list but only 45 of those are $K3 \times K3$ configurations. 

\begin{figure}[t]\centering
\includegraphics[width=0.85\textwidth]{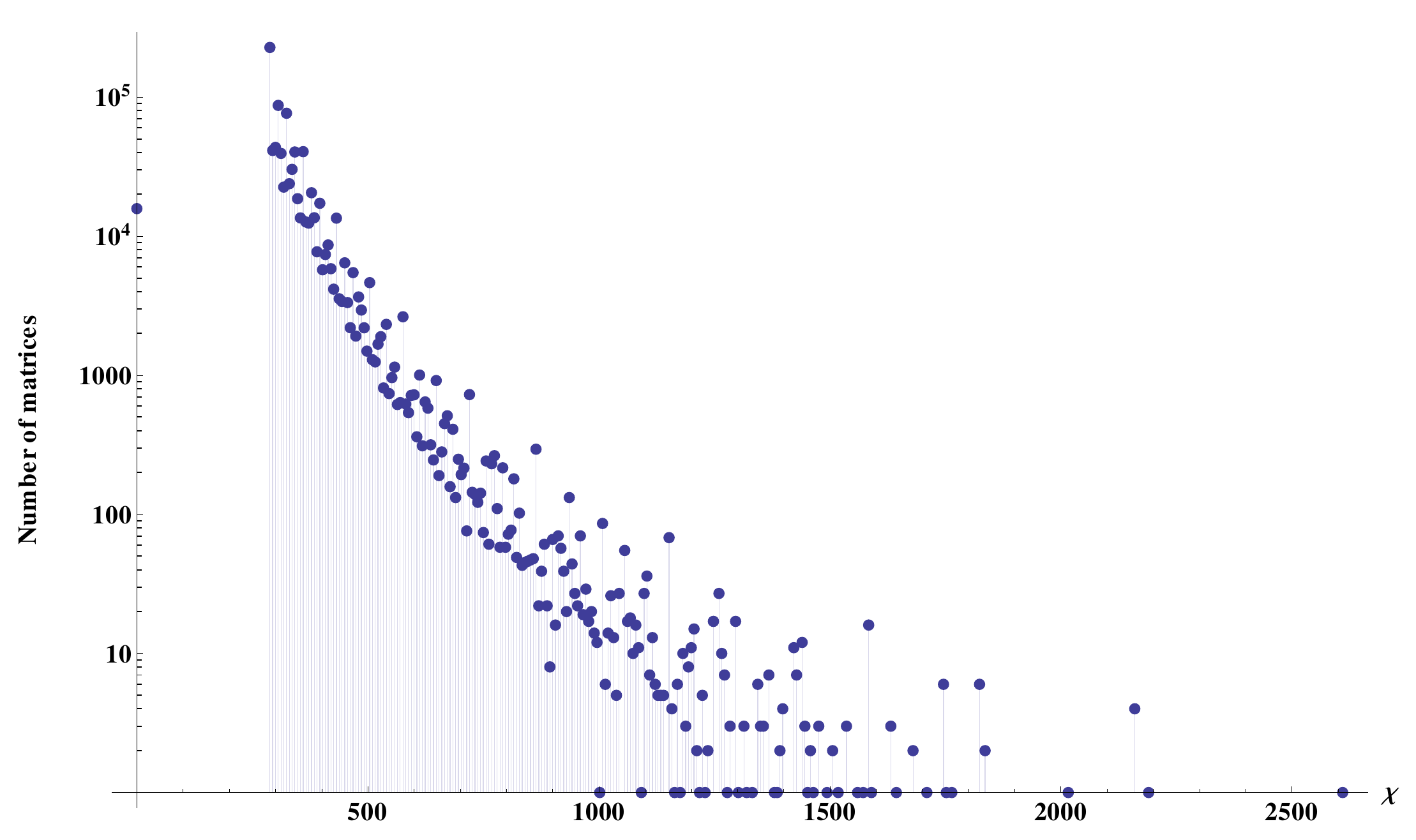}
\caption{Distribution of the Euler characteristic $\chi$ in the CICY four-fold list, as a logarithmic plot. The values lie in the range $0\leq \chi \leq 2610$.}
\figlabel{eulerhisto}
\end{figure}
The Euler characteristic for each of the 921,497 matrices was computed and found to be in the range $0\leq \chi \leq 2610$. As mentioned above, all configurations with Euler characteristic 0 correspond to direct product manifolds and the non-zero values for the Euler characteristic are found to be in the range $288\leq \chi\leq 2610$. A logarithmic plot of the distribution of Euler characteristics is shown in \figref{eulerhisto}. About 25\% of all matrices have Euler characteristic equal to 288, the smallest non-zero value in the list. This huge peak at a single value might indicate non-trivial residual redundancies in the list. The full list of configuration matrices with Euler characteristics can be downloaded from~\cite{cicylist4}.

In total, the list contains 206 different values of $\chi$ and, hence, this provides a weak lower bound on the number of inequivalent CICY four-folds. As already mentioned, this bound can be significantly strengthened by computing additional topological data, such as Hodge numbers, Chern classes and intersection numbers. A detailed analysis will be presented in ref.~\cite{paper2}, but a preliminary calculation shows that the data set contains at least 3737 different sets of Hodge numbers. Computing even finer topological invariants will strengthen this bound further.

Finally, we should address the question of how many CICY four-folds in our list have an elliptic fibration structure. We will not attempt to answer this question in full generality since a necessary and sufficient criterion for the existence of such an elliptic fibration which is suitable for practical computations is currently not known. Fortunately, for CICYs there is a particularly simple type of elliptic fibration which is consistent with the embedding in the projective ambient space. Suppose a configuration matrix $[{\bf n}|{\bf q}]$ for a CICY four-fold can be brought, by a combination of row and column permutations, into the equivalent form
\be\eqlabel{fibconf}
 \left[\begin{array}{c|cc}{\bf n}_F&F&{\bf 0}\\{\bf n}_B&C&B\end{array}\right]\; ,
\ee
such that the sub-configuration $[{\bf n}_F|F]$ is a one-fold. Then, the CICY four-fold is elliptically fibered with $[{\bf n}_F|F]$ representing the $T^2$ fiber and $[{\bf n}_B|B]$ the three-fold base while the entries $C$ describe the structure of the fibration, that is, the way in which the fiber is twisted over the base.

We have checked how many CICY configuration matrices from our list can be brought into the form~\eqref*{fibconf}. It turns out that this is possible for all but $477$ of the 921,497 matrices. Moreover, in many cases a given configuration matrix can be brought into the form~\eqref*{fibconf} in many different, inequivalent ways, indicating the existence of inequivalent fibrations. Unfortunately, an elliptic fibration structure of this kind does not automatically imply the existence of a section. However, a preliminary analysis shows that the vast majority of manifolds indeed admit fibrations which do have sections. Details of this analysis will be presented in ref.~\cite{paper2}.

\section{Summary and outlook}\seclabel{outlook}

In this paper, we have classified all complete intersection Calabi-Yau four-folds (CICYs) in ambient spaces which consist of products of projective spaces. We have found a list of 921,497 configuration matrices which represent all topologically distinct CICYs. This is to be compared with 7890 configuration matrices which were found in the analogous classification for CICY three-folds carried out in ref.~\cite{Candelas:1987kf}. A total of 15813 configuration matrices from our four-fold list describe direct product manifolds of various types but all other matrices represent non-decomposable CICY four-folds. Discarding the cases with Euler characteristic $0$ which all correspond to direct product manifolds, the Euler characteristic is in the range $288\leq \chi\leq 2610$. The list contains 206 different values for the Euler characteristic, a weak lower bound for the number of inequivalent CICY four-folds. This bound can be strengthened by considering additional topological invariants. For example, a preliminary analysis shows that the list contains at least 3737 different sets of Hodge numbers. We have also studied the existence of a particular class of elliptic fibrations, consistent with the projective embedding of the manifolds, and have found that almost all manifolds in our list are elliptically fibered in this way. Often, a given CICY four-fold allows for many fibrations of this kind. A preliminary analysis shows that most of these manifolds admit such fibrations which have sections. 

We hope that the data set compiled in this paper will be of use in various branches of mathematics and physics. Due to their embedding in projective ambient spaces, CICYs are particularly simple and many of their properties are accessible through direct calculation. In the context of string theory, Calabi-Yau four-folds can be used for string compactifications, for example of type II or heterotic theories to two dimensions or, perhaps most importantly, of F-theory to four dimensions. F-theory compactifications require elliptically fibered Calabi-Yau four-folds, preferably with a section, and we have seen that our manifolds support these properties. 

We have left a number of more advanced issues for a longer companion paper~\cite{paper2} which is currently in preparation. These include the calculation of Hodge numbers, Chern classes and intersection numbers as well as a more detailed analysis of elliptic fibrations. This additional data will allow us to place a more realistic lower bound on the number of inequivalent CICY four-folds. It will also facilitate applications, particularly in the context of F-theory.

\acknowledgments

The authors are very grateful to Yang-Hui He for collaboration in the early stages of this work. We would also like to thank Philip Candelas, Kelly Stelle and David Weir for discussions. The work of J.~G.~was partially supported by NSF grant CCF-1048082, CiC (SEA-EAGER): A String Cartography. A.~L.~is partially supported by the EC 6th Framework Programme MRTN-CT-2004-503369 and by the EPSRC network grant EP/l02784X/1.


\end{document}